\documentstyle[12pt]{article}
\begin{document}

  \begin{center}
{\bf      S-wave $\pi$-Nucleus Repulsion and Dirac Phenomenology}\vspace{2cm}\\
S. Chakravarti\\
{\em Department of Physics, California State Polytechnic University,
Pomona, CA 91768, USA}\vspace{1cm}\\
B.K. Jennings\\
{\em TRIUMF, 4004 Wesbrook Mall, Vancouver, B.C. V6T 2A3, Canada}\vspace{2cm}\\
  \end{center}

\vspace{1cm}
\begin{center}
  Abstract
\end{center}

\vspace{1cm}
A relativistic $\pi$-nucleon potential is extended to $m^* \neq m$ to
investigate the possibility of generating s-wave $\pi$-nucleus repulsion. We
find that relativity does indeed generate significant repulsion, the exact
amount depending on the details of the calculation. In contradistinction the
$t\rho$ approximation gives very little repulsion.

  \vspace{1cm}
  \begin{center}
    Submitted to Physics Letters\\
\today
  \end{center}

\pagebreak

The experimental data on low-energy pion-nuclear
interactions \cite{kluge} from a variety of sources, e.g., pionic atoms at
zero pion kinetic energy
$E_{\mbox{kin}}$,  elastic, inelastic and charge exchange reactions at
$E_{\mbox{kin}}  \le$
50 MeV, etc., are generally well described by theoretically motivated
phenomenological optical potentials \cite{pinucleus}. These potentials are
generated by using the nonrelativistic impulse approximation plus density
dependent corrections.  Unfortunately they fail to produce
the central pion-nucleus repulsion of about $30$ MeV required by fits of the
experimental data \cite{olivier}.

The underlying isoscalar $\pi$N s-wave interaction is
exceptionally small at threshold (with a scattering length $b_0 =
-0.010~m_{\pi}^{-1}$, $m_{\pi}$ being the pion mass), its double-scattering
modification \cite{hufner} is relatively ineffective, and the dispersive
contribution accompanying the imaginary $\pi$NN absorption term is believed to
be attractive \cite{oset}; if this latter contribution were to blame, it would
have to be {\it repulsive}, with a magnitude of up to five times its imaginary
counterpart.

Recently there has been considerable controversy over the role of relativity in
addressing this problem.  Birbrair and Gridnev \cite{birbraira,birbrairb} have
made the observation that, to the extent that Nuclear Dirac Phenomenology
provides a valid and meaningful description of the gross properties of nuclei
at low energies, it naturally offers additional s-wave repulsion.

This approach has been attacked from two fronts. First, it has been shown that
for pion scattering there are large off-shell ambiguities \cite{gal}. Second,
it has been argued that for pions the impulse approximation is not valid, since
using $m^*$ internally in diagrams cancels most of the $m^*$ effects coming
from the impulse approximation \cite{koch,osetb}. These are both valid
concerns.

One way to address these concerns is to take a relativistic model that
describes  the free $\pi$-nucleon scattering and
examine its prediction when the nucleon effective mass $m^*$ is varied.
The off-shell behavior is then fixed by the model and is no longer ambiguous.
In addition $m^*$ is included in all possible places and one goes beyond the
impulse approximation.  In this letter we use the relativistic $\pi$-nucleon
potential developed by Pearce and Jennings \cite{pearce} to carry out this
program.  We find that there is a repulsion on the order of a few tens of MeV
coming from the relativistic effects. However,
we also find that both objections to
the work of Birbrair and Gridnev are correct.
The $t\rho$ approximation gives very little dependence on $m^*$
while a significant effect arises from
putting $m^*$ in intermediate propagators.

The present work is a first step towards developing a full relativistic optical
potential. That work would require the consideration of other partial waves and
the treatment of Pauli blocking. A relativistic potential is interesting in its
own right and, as pointed out in ref.~\cite{hicks}, is crucial in reactions
like (p,n$\pi^+$) where the nucleons are treated most easily in a Dirac
approach.

An early attempt to develop a relativistic model was made by Miller and
Noble \cite{miller}. They used the $\sigma$-$\omega$ model that did not attempt
to simultaneously describe the free nucleon-nucleon scattering. Thus it is
quite
different from our approach.

An alternate approach to the relativistic pion-nuclear problem has been
developed by Leisi and co-workers \cite{leisi}. This approach introduces an
explicit scalar field ($\sigma$) for the pion to couple to. In our work we are
only concerned with the effects of the pion coupling to an in-medium nucleon
and have nothing to say, at the present time, about an explicitly enhanced
$\sigma$ field.

We start with the $\pi$-nucleon potential developed by Pearce and Jennings
\cite{pearce}. That potential includes $\rho$, $\sigma$ exchange and the
nucleon and delta pole and crossed-pole terms. The nucleons are treated as
Dirac particles and
negative energy propagators are included at each step of the process. In the
present work we simply take the Pearce-Jennings model and replace $m$ with
$m^*$. In
particular the mass appears in the energy denominators of the nucleon and delta
pole and crossed-pole terms and in the free spinors. One has to be careful not
to replace $m$ where it has a purely formal role such as in the definition of
the pseudo-vector coupling.  We have used the smooth propagator
\cite{jennings} and have
checked that our results agree with ref.~\cite{pearce} when $m^*=m$.

The calculation of phase shifts uses bare values for the nucleon and delta
masses and couplings of the pole diagrams in the $P_{11}$ and $P_{33}$
channels. These are calculated by the renormalisation procedure described in
ref.~\cite{afnan} wherein the $P_{11}$ amplitude is required to have a pole at
the physical nucleon mass $m$, and the residue is required to reproduce the
physical $\pi$NN coupling constant. The bare delta mass and coupling constant
are treated as parameters and adjusted to fit the experimental phase shifts.

The values to use for the in-medium $\pi$NN and $\pi$N$\Delta$ couplings are
ambiguous.  The obvious thing to do is to keep the bare couplings fixed.  The
renomalisation will then occur due to the dressing generated by solving the
Lippmann-Schwinger equation. This will work well in the p-waves where the
renormalisation is generated directly in this manner in free space.  In this
case the physical couplings will vary with the effective mass $m^*$.

For the s-waves the physical couplings are put in by hand rather than being
generated within the model. It is not clear how they will change in the medium.
A possible constraint arises from the Goldberger-Treiman relation. It has been
suggested that in-medium we have two Goldberger-Treiman relations \cite{riska}
depending on which quantum numbers we are considering.  For the s-wave the
renormalisation is different from that for the p-waves and is quite
small \cite{riska}. This suggests that one keep the couplings fixed at their
physical values. We will give results for both cases.

In our work we chose $m^*$ as the (variable) physical nucleon mass and obtained
the corresponding bare nucleon mass by the renormalisation procedure.  If we
keep the bare coupling constants fixed at the value for $m^{*}=m$ the
renomalised values vary with $m^*$. The values of the dressed coupling
constant and the bare nucleon mass obtained this way are shown in table 1.

To obtain the dressed delta parameters as a function of $m^*$, we set the
dressed delta-nucleon mass difference $m^{*}_\Delta$-$m^*$ equal to the {\it
physical} mass difference.  Calculations were then performed for the $P_{33}$
channel at energies around resonance ($\delta_{P_{33}}$ = 90$^o$). The value of
the dressed delta mass was then extracted from the location of the resonance,
and that of the dressed coupling constant $f_{\pi N \Delta}$ from the slope of
the phase shift curve as a function of energy. Values of the extracted
parameters are shown in table 2.

The results for the scattering lengths in each isospin channel are given in
table 3 for three cases:
fixed bare couplings, fixed physical couplings and the $t\rho$ approximation.
For each case, we also give the isoscalar pseudopotentials \cite{weise}
obtained from the scattering lengths,
using a nuclear matter density of 0.17 fm$^{-3}$.

It is instructive to examine, for a fixed pion energy, the contributions to
the on-shell Born term \cite{pearce} from the individual diagrams in each
isospin channel as a function of $m^*$ . These are shown in figs. 1 and 2
for the case of fixed bare couplings, with $T_{\pi}$ = 20 MeV.
%
%The contributions to the pseudopotential from the individual diagrams
%in each isospin channel are shown in figs. 1 and 2.
%
We see that the main
variation is in the pole and crossed-pole terms. The variation has two main
sources.  First, the energy denominator varies approximately as $1/m^*$.
Second, the couplings vary (when we keep the bare couplings fixed).

Next, we note that in the $t\rho$ approximation there is very little $m^*$
dependence. This is consistent with the off-shell extrapolation used by Gal
{\it et al.} \cite{gal} but inconsistent with the extrapolation of Birbrair
{\it et al.} \cite{birbraira,birbrairb}. Thus we conclude that the $t\rho$
approximation cannot account for the repulsion.

We see, however, that including $m^*$ also in the interior of diagrams does
give repulsion. This comes about from the energy denominators of the z-graphs.
The most repulsion arises when we keep the bare coupling constants fixed and
gives about 20 MeV repulsion at nuclear matter density.  When the dressed
couplings are kept fixed at their physical values (which we believe is more
realistic) the repulsion is cut almost in half. Thus we do not have enough
repulsion to cure the entire problem but the effect is in the correct direction
and within a factor of 2 of the correct size.

In summary, we have examined the predictions of a relativistic $\pi$-nucleon
potential when $m^* \neq m$ for two different choices of the in-medium
coupling constants, and compared them with that of the $t\rho$ approximation.
We have found that relativity does indeed generate significant repulsion,
the exact amount depending on the choice of the in-medium couplings. In
contradistinction the $t\rho$ approximation is found to give very little
repulsion.

S.C. gratefully acknowledges the hospitality of the TRIUMF Theory group
where this work was begun. The authors acknowledge the hospitality of the
National Institute of Nuclear Theory, University of Washington, where part
of this work was carried out.  B.K.J. acknowledges financial support from the
Natural Sciences and Engineering Council of Canada.

\newpage
\begin{center}
Figure Captions
\end{center}

\vspace{1cm}
Fig.~1. The on-shell Born potential in the S11 channel (solid line) as a
function of $m^*/m$, for $T_{\pi}$ = 20 MeV and fixed bare couplings.
The contributions from the individual diagrams are labelled as follows:
(a): nucleon pole; (b): crossed nucleon pole; (c): $\rho$ exchange;
(d): $\Delta$ pole; (e): crossed $\Delta$ pole;
(f): $\sigma$ exchange.

\vspace{0.6cm}
Fig.~2. Same as in fig. 1 but for the S31 channel.

\newpage
\begin{center}
Table 1. The values of the dressed $\pi$NN coupling constant and the bare
nucleon mass for a range of $m^*$.

\vspace{.5cm}
\begin{tabular}{|ccc|}\hline
$m^{*}/m$ & $g_{\pi NN}^2/(4\pi)$ &  $m_{oN}$ \\
       &              &  (MeV)  \\ \hline
0.5    &   19.85      &   652.4 \\
0.6    &   19.07      &   739.9 \\
0.7    &   18.01      &   827.7 \\
0.8    &   16.79      &   915.8 \\
0.9    &   15.53      &  1004.3 \\
1.0    &   14.30      &  1093.2 \\
1.1    &   13.16      &  1182.7 \\
1.2    &   12.14      &  1272.7 \\ \hline
\end{tabular}
\end{center}

\vspace{.5cm}
\begin{center}
Table 2. Values of the dressed $\pi$N$\Delta$ coupling constant and $\Delta$
mass for a range of $m^*$.

\vspace{.5cm}
\begin{tabular}{|cccc|}\hline
$m^{*}/m$ & $f^{2}/(4\pi)$ & $m_{\Delta}^{*}$ & $\Gamma_{\Delta}$\\
        &           &  (MeV)     & (MeV) \\\hline
0.5     &  0.496    &     826.2  &  169 \\
0.6     &  0.538    &     893.1  &  173 \\
0.7     &  0.522    &     968.8  &  161 \\
0.8     &  0.475    &    1052.3  &  144 \\
0.9     &  0.415    &    1140.7  &  128 \\
1.0     &  0.365    &    1231.8  &  113 \\
1.1     &  0.308    &    1324.4  &  102 \\
1.2     &  0.270    &    1417.6  &   92 \\ \hline
\end{tabular}
\end{center}
\newpage
\begin{center}
Table 3. Scattering lengths and pseudopotential strengths for a range of $m^*$.

\vspace{.5cm}
\begin{tabular}{|c|ccc|ccc|ccc|}\hline
&\multicolumn{3}{c|}{In Medium}&\multicolumn{3}{c|}{Free}&
\multicolumn{3}{c|}{$t\rho$}\\
&\multicolumn{3}{c|}{Couplings}&\multicolumn{3}{c|}{Couplings}&
\multicolumn{3}{c|}{Approximation}\\\hline
$m^*/m$&$a_{\rm S11}$&$a_{\rm S31}$&$V$&$a_{\rm S11}$&$a_{\rm S31}$&$V$&$a_{\rm
  S11}$&$a_{\rm S31}$&$V$ \\
     &(fm) &(fm) &(MeV) &(fm) &(fm)  &(MeV) &(fm)  &(fm)  &(MeV) \\  \hline
0.5& 0.108& -0.184&  26.2& 0.134& -0.146&  16.0& 0.226& -0.123&   2.1\\
0.6& 0.118& -0.183&  25.0& 0.150& -0.142&  13.4& 0.222& -0.121&   2.0\\
0.7& 0.140& -0.169&  20.1& 0.167& -0.136&  10.6& 0.218& -0.119&   2.0\\
0.8& 0.166& -0.151&  13.8& 0.183& -0.129&   7.6& 0.216& -0.118&   2.0\\
0.9& 0.191& -0.132&   7.3& 0.199& -0.122&   4.7& 0.214& -0.117&   2.0\\
1.0& 0.213& -0.116&   2.0& 0.213& -0.116&   2.0& 0.213& -0.116&   2.0\\
1.1& 0.230& -0.103&  -2.4& 0.225& -0.110&  -0.5& 0.211& -0.115&   1.9\\
1.2& 0.243& -0.093&  -5.7& 0.235& -0.105&  -2.6& 0.210& -0.115&   1.9\\
\hline
\end{tabular}
\end{center}
\end{document}